\documentclass[a4paper,11pt]{article}
\usepackage{etoolbox}
\usepackage{jheppub} 
\usepackage{braket}

\usepackage[T1]{fontenc} 
\usepackage{braket}
\usepackage{xspace}
\usepackage{amsmath}
\usepackage{amsfonts}
\usepackage{amssymb}
\usepackage{lineno}

\title{ \boldmath Group averaging and BRST quantization in de Sitter space}

\author[a]{Mendrit Ljatifi}

\affiliation[a]{Institute for Theoretical Physics, Heidelberg University,\\Philosophenweg 19, Germany}

\emailAdd{ljatifi@thphys.uni-heidelberg.de}

\abstract{In this note we give a BRST interpretation of inner product of group averaging on de Sitter group $SO(2,1)$. Quantization approaches can be divided in two classes, without introducing additional degrees of freedom and with introducing ghosts, antighosts, Lagrange multipliers and canonically conjugated momenta. The first class of quantization methods includes Dirac method and group averaging method. The Dirac method is based on imposing the constraint conditions on physical states, while in group averaging method one modifies the physical inner product due to constrains.\\
One should distinguish the abelian case for which you can have continuous spectrum and discrete spectrum for which there are internal topological problems in BRST-BFV approach, and the non- abelian case.}

\begin{document} 
\maketitle
\flushbottom
\label{sec:intro}
\section{Introduction}
In this article, we review Refined Algebraic Quantization(RAQ) and the group averaging on a de Sitter group and give a possible correspondence with the BRST quantization.\\
Refined Algebraic Quantization is a set up for implementing Dirac quantization method by beginning with unconstrained system in which also the gauge depended operators act on the "primary" auxiliary Hilbert space $\mathcal{H}_{aux}$.
One can introduce constraints as operators $C_i$ on some space and then take states that only are annihilated by those constraints which we label as physical states $\ket{\psi}_{phys}$ \cite{Higuchi:1991tk,Higuchi:1991tm,Giulini:1998kf}. An important step in RAQ is defining a subspace $\Phi \in \mathcal{H}_{aux}$ which is mapped into itself by constraints, that is  $C_i: \Phi \rightarrow \Phi$. Since RAQ looks for solutions of the constraints in $\Phi^*$ which is the algebraic dual of $\Phi$. For RAQ to be complete we need a map, say $\eta$, from $\Phi$ to $\Phi^*$, $\eta : \Phi \rightarrow \Phi$ called the "rigging" map. The importance of the "rigging" map is that we can write the Hermitian inner product on the image of $\eta$ completing into a Hilbert space \cite{Higuchi:1991tk,Higuchi:1991tm,Giulini:1998kf,Giulini:1998rk,Marolf_2009}:
\begin{equation}
    \big( \eta(\psi_1),\eta(\psi_2) \big)_{phys} = \eta(\psi_1)[\psi_2]
\end{equation}
The following of the article is constructed as follow; in Section 2, we review group averaging method for a general theory, in Section 3 we review free scalar fields in (1+1) dimensional de Sitter space. In Section 4 we calculate the inner product of group averaging method for $SO(2,1)$ group while in Section 5 we give the BRST interpretation of the inner product \eqref{innerprod}.\\
Finally in Section 6, following \cite{Witten:2022xxp} we discuss a possible gauge fixing model that can be constructed for our interpretation.

\label{sec:RGA}
\section{Review of Group Averaging}
A powerful technique for studying constrained systems is what is called group averaging \cite{Higuchi:1991tk,Higuchi:1991tm}, although it is likely to be well defined except in certain limited cases it still allows us to many constrained systems and physical theories.\\
Recall that Dirac method of quantization involves introducing the constrains as operators on some space and then taking those states which are annihilated by the constrains to be what we call physical states \cite{Higuchi:1991tm,Higuchi:1991tk}.Considering that Dirac procedure is closely related to BRST approach \cite{Becchi:1974md} those become very powerful and useful techniques when studying quantum gauge systems.\\
There are number of variants of Dirac method that have been discussed \cite{Woodhouse:1981zd} and also \cite{Higuchi:1991tk,Higuchi:1991tm,Kuchar:1986jj,Ordonez:1995kz} and Refined algebraic quantization(RAQ)\cite{Higuchi:1991tk,Higuchi:1991tm,Ashtekar:1995zh,Giulini:1998rk}.\\
However RAQ technique becomes much more powerful when is possible to apply what is known as "group averaging". Group averaging starts with the integral
\begin{equation} \label{firstintegral}
    \int_G \braket{\phi_1| U(g) | \phi_2} dg
\end{equation}
over the gauge group $G$ where we define the physical Hilbert space and $dg$ is the Haar measure on G \cite{Giulini:1998kf}. In particular, group averaging provides the unique Hilbert space representation with a unique inner product of the algebra of observables which is compatible with RAQ \cite{Gomberoff:1998ms,Chandrasekaran:2022cip,Chandrasekaran:2022eqq}. As described in \cite{Giulini:1998kf} the convergent group averaging ensures a unique representation of the algebra of observables and it must diverge in the presence of any superselection rules \cite{Giulini:1998kf}.\\
In the following, we consider the group $G=SO_n(n,1)$ acting on $L^2(M^{n,1},d^n x)$ where $L^2$ is the set of square integrable $L^2$-functions. The infinitesimal action of group is defined by generators of the Lie Algebra, 
\begin{equation}
    J_{\mu \nu}= \eta_{\mu \nu} x^\alpha  \frac{\partial}{\partial x^\nu} - \eta_{\nu \alpha} x^\alpha \frac{\partial}{\partial x^\alpha}
\end{equation}
taking the exponent of generators  gives the unitary action $U(g)$ on the group. Generators $J_{\mu \nu}$ also define the constrains of the theory where physical states satisfy the condition 
\begin{equation}
    J_{\mu \nu} \ket{\psi}_{phys}=0
\end{equation}
and since there is no such a normalizable states, in the language of  REQ this condition can be written as 
\begin{equation}
    \bra{\psi}_{phys} J_{\mu \nu}=0
\end{equation}
When computing \eqref{firstintegral} we are interested in some domain $\Phi$, if the integral converges or if it can be defined in a sensible way, it will define a map (which we know as "rigging map") from $\Phi$ into the space of physical states. Before going into the calculation, we first need to compute its Haar measure. Any element $g$ in $G=SO_c(n,1)$ is a product of a boost and a rotation and we can use Cartan decomposition \cite{Barut:1986dd}. Choosing $x_0$ as time coordinate in Minkowski space, we can write an element of $G$ as $ g=h_0 k_0$ for $k_0 \in K=SO_c(n)$ subgroup of $G$ that preserves the $x_0$ axis and $h_0$ a symmetric positive definite matrix(boost). In general for $h_0$ we can write $h_0=k_1 b(\lambda) k_1^{-1}$ for a rotation $k_1 \in K$ and $b(\lambda)$ a boost with $\lambda$ being the boost parameter in $x^0, x^1$-plane.\\
It is convenient to write $k=k_1^{-1} k_0$ and $h=k_1 b(\lambda)$ so that we have
\begin{equation}
    g=hk, \, k\in K, \, \, \, \text{and} \quad   h \in H_+^{n}
\end{equation}
where $H_+^{n}$ can be identified as the right coset space $SO_c(n,1)/K$. With a help from the equation of hyperboloid, we can represent this space as the upper sheet of it hyperboloid 
\begin{equation} \label{hyperbolaeq}
    -(x^0)^2+ (x^1)^2+...+(x^n)^2=-1
\end{equation}
A generic element of $H_+^n$ can be written as
\begin{equation} \label{genericelement}
    h= k_{n-1}(\theta_{n-1})k_{n-2}(\theta_{n-2}) ... k_1(\theta_1) b(\lambda)
\end{equation}
where $k_m$ is a rotation in the plane $(x^m,x^{m+1})$ and $b(\lambda)$ is a hyperbolic rotation in $(x^0,x^1)$. \\
Using \eqref{hyperbolaeq} and \eqref{genericelement}, in Minkowski coordinates we can write
\begin{equation}
    \begin{split}
        x^0 &= \cosh{\lambda} \\
        x^1 &= \sinh{\lambda} \cos{\theta_1} \\
        x^2 &= \sinh{\lambda} \sin{\theta_1} \cos{\theta_2} \\
        \vdots\\
        x^n &= \sinh{\lambda} \sin{\theta_1} \sin{\theta_2} ... \sin{\theta_{n-2}} \sin{\theta_{n-1}} \\
    \end{split}
\end{equation}
Writing the measure $d^{n+1}x$ as $s^n ds \, dh$ leads to an $SO_c(n,1)$-invariant measure $dh$ for $H_+^n$ given by 
\begin{equation}
    dh = \sinh^{n-1}{\lambda} \sin^{n-2}{\theta_1}... \sin{\theta_{n-2}} d \lambda d \theta_1 ... d \theta_{n-1}
\end{equation}
where $d \lambda$ and $d \theta_i$ are the Lebesgue measures on certain intervals. Following \cite{Gomberoff:1998ms,Ashtekar:1995zh} we can write $dg$ as $dg =dh dk$ as a Haar measure on G.\\
Once we have the Haar measure, we can proceed with calculation of \eqref{firstintegral}. We want to study this integral in some domain $\Phi \in \mathcal{H}_{aux}$. We proceed by introducing distributional states $\ket{x}$ for $x \in M^{n,1}$ which satisfy 
\begin{equation}
\braket{x_1 | x_2}= \delta^{n+1}(x+1,x_2)
\end{equation}
where now we can study \eqref{firstintegral} as a distribution in both $x_1$ and $x_2$, for which we can write
\begin{equation} \label{integral1}
\begin{split}
   I= \int dg \braket{x_1 |U(g)|x_2} =& \frac{1}{V_{SO(n)}} \int dk \int dg \braket{x_1 |U(kg)|x_2} \\
    =& \frac{1}{V_{SO(n)}} \int dk dh dk^\prime \braket{x_1 | U(khk^\prime)|x_2}
    \end{split}
\end{equation}
where $V_{SO(n)} = \int dk$ is the volume of $SO(n)$. Using the $SO(n)$ translation invariance of $dk$ \eqref{integral1} can be written as
\begin{equation}
    I=\frac{V_{S_{n-1}}}{V_{SO(n)}} \int dk dk^\prime d \lambda \sinh^{n-1}{\lambda} \braket{x_1 |U(k)U(b(\lambda)) U(k^\prime) |x_2}
\end{equation}
where $V_{S_{n-1}}= \frac{\pi^{n/2}}{\Gamma(n/2)}$ is the volume of the $(n-1)$-sphere. Following \cite{Gomberoff:1998ms} the integral I can be easily done for $n=1$,
\begin{equation}
    I_{n=1}= \delta \big( (x_1^2-t_1^2),(x_2^2-t_2^2) \big)
\end{equation}
This result is in analogy with the integral $\int e^{iks} dx$ which converges to $\delta(k)$.\\
For $n>1$ the integral $I$ can be written as
\begin{equation}
    I = \frac{\delta(s_1^2,s_2^2)}{r_1^{n-2} r_2^{n-2}} \int d \xi \Big[ s_1^2 \xi^2 +2 t_1 t_2 \xi -(r_1^2+t_2^2)   \Big]^{\frac{n-3}{2}}
\end{equation}
where $s_a^2 = \eta_{\mu \nu} x_a^\mu x_a^\nu$, for $a=1,2$, $r^2= \sum_{i>0} x^i x^i$, and $\xi=\sinh{\lambda}$.\\
In the case when $x_1$ is time-like  and $x_2$ is space-like  the integral \eqref{firstintegral} will vanish, while in the case when both $s_1$ and $s_2$ are space-like the situation changes. In that case one can write $I$ as a divergent factor times a Lorentz invariant quantity \cite{Gomberoff:1998ms} while for $s_1$ and $s_2$ it can be shown \cite{Gomberoff:1998ms} that the integral  \eqref{firstintegral} is convergent for $n>1$. 

\section{Free scalar fields in 1+1 de Sitter}

Any theory of quantum gravity should include a description of de Sitter space \cite{Marolf_2009,Hajicek:1994px,Higuchi:1991tk}, at least as an approximation. We recall that field theories on spacetimes with Killing symmetries have conserved charges and we regard this type of theory as zero-order perturbative approximation to a theory of matter plus gravity. This is interesting in the case when the background also has a compact Cauchy surface, then if you apply the equivalent of Gauss' law for gravity, the above charge must vanish in order for a solution to consistently couple to dynamical gravity. These are known as linearization-stability constrains and require that in de Sitter space, linearized  quantum states to be invariant under the de Sitter group $SO_0(D,1)$ where $D$ is the space-time dimension.\\
Because the de Sitter group is non-compact, the standard Fock which we call  $\mathcal{H}_{aux}$ space contains no de Sitter-invariant states except for a possible vacuum. This however is not a rich physical theory and we need a better description of the Fock space. One may use the standard Fock space to build a new physical one that we call $\mathcal{H}_{phys}$. We shall call this physical Hilbert space of de Sitter invariant states. The way we build the new physical space is via group averaging, where one considers linear superposition of auxilary states $|\ket{\psi}$ of the form 
\begin{equation} \label{state1}
    \ket{\psi} : = \int_{g \in G} dg U(g) |\ket{\psi}
\end{equation}
where G is the de Sitter group, $dg$ is the unique Haar measure of G, and $U(g)$ gives the unitary representation of G on $\mathcal{H}_{aux}$. Since the de Sitter group G is non-compact, the state \eqref{state1} is not normalizable in $\mathcal{H}_{aux}$. However we can define a new inner product on the group averaged states as 
\begin{equation}
    \bra{\psi_1} \ket{\psi_2}:= \bra{\psi_1} | \cdot | \ket{\psi_2} = \int_{g \in G} dg \bra{\psi_1} || U(g) || \ket{\psi_2}
\end{equation}
In this section we review the group averaging for massless scalar field in $1+1$ de Sitter space. We write the $1+1$ de Sitter metric as 
\begin{equation}
    ds^2 = \frac{l^2}{\cos^2 \tau} (-d\tau^2 + d\theta^2)
\end{equation}
This form of metric is adopted  from conventions of conformal field theory which is just a conformal factor times the metric on the cylinder. $\tau$ is the conformal time and has a range $ -\pi/2 < \tau <  \pi/2$, and $\theta$ is periodic $\theta \equiv \theta + 2 \pi$ and $l$ is the de Sitter length scale.  Adopting the light-cone coordinates, the metric can be written as 
\begin{equation}
    ds^2 =\frac{l^2}{2 \cos^2 \tau} \big( (dx^+)^2 + (dx^-)^2 \big)
\end{equation}
The action of a free scalar field is 
\begin{equation}
    S= -\frac{1}{2} \int dx^2 \sqrt{-g} g^{ab} \nabla_a \phi \nabla_b \phi = \int dx \partial_+ \phi \partial_- \phi
\end{equation}
The equation of motion for $\phi$ is given by $\partial_+ \partial_- \phi =0 $ where the solutions can be decomposed into left- and right-moving modes as
\begin{equation}
    \partial_+ \phi(x^+)= \frac{1}{2\sqrt{\pi}} \sum_m \alpha_m \exp[-imx^+], \quad \quad \partial_- \phi(x^-) =\frac{1}{2 \sqrt{\pi}} \sum_m \Tilde{\alpha}_m \exp[-imx^-]
\end{equation}
Integrating and combining the solution we find that 
\begin{equation}
    \phi(x)= \frac{\phi_0}{4 \pi}+ \alpha_0 x^+ + \Tilde{\alpha}_0 x^- + \frac{i}{2 \sqrt{\pi}} \sum_{m \neq 0} \Big[ \frac{\alpha_m}{m} e^{-imx^+}+ \frac{\Tilde{\alpha}_m}{m} e^{-imx^-}\Big]
\end{equation}
substituting back the light-cone coordinates, $x^{\pm}= \tau \pm \theta$ we get
\begin{equation} \label{phisol}
    \begin{split}
        \phi(x) = & \frac{\phi_0}{4 \pi} + (\alpha_0 + \Tilde{\alpha}_0) \tau + (\alpha_0  +\Tilde{\alpha}_0) \theta \\
        & + \frac{i}{m \sqrt{\pi}} \sum_{m \neq 0} \frac{1}{m} e^{-im \tau} \Big[ \alpha_m e^{-im\theta} +\Tilde{\alpha}_m e^{+im \theta} \big]
    \end{split}
\end{equation}
The term linear in $\tau$ we label it as a momentum $ p  \varpropto (\alpha_0 - \Tilde{\alpha}_0)$. Since $\phi(x)$ must be single-valued, we can consider two cases.\\
1. The target space of $\phi(x)$ is the real line and if we impose the periodicity condition of $\theta$, that is $\phi(\tau,\theta) = \phi(\tau,\theta + 2 \pi)$ we get $\alpha_0 = \Tilde{\alpha}_0$ and the linear term in $\theta$ in \eqref{phisol} vanishes.\\
2. The second case would be if target space of $\phi(x)$ is a circle $S^1$ with radius $R$. Since $\phi(x)$ must be single valued, this requires that 
$\phi(\tau, \theta + 2 \pi)= \phi(\tau,\theta)+ 2 \pi R  w$ where $w$ is the winding number of the field. From \eqref{phisol} we can see that $Rw= (\alpha_0, \Tilde{\alpha}_0)$.\\
Now, if we combine both these cases, since $\phi(x)$ is periodic, then $p$ is quantized: $p=k/R$, for $K \in \mathbb{Z}$. We can write our mode expansion \eqref{phisol} as 
\begin{equation}
    \phi(x) = \frac{\phi_0}{4 \pi} + 2 p\tau + Rw \theta + \frac{i}{m \sqrt{\pi}} \sum_{m \neq 0} \frac{1}{m} e^{-im \tau} \Big[ \alpha_m e^{-im\theta} +\Tilde{\alpha}_m e^{+im \theta} \big]
\end{equation}
We continue the quantization of scalar field using the canonical techniques where at the end we get the auxilary Hilbet space $\mathcal{H}_{aux}$. In canonical quantization, $\phi_0, p, w, \alpha_m,$ and $ \Tilde{\alpha}_m$ become operators. Imposing the canonical commutation relation $ [\phi(\tau,\theta_1),\phi(\tau,\theta_2)]= i \delta(\theta_1 -\theta_2)$ one finds that 
\begin{equation}
    [\phi_0, p]=i, \quad [\alpha_m,\alpha_n] =[\Tilde{\alpha}_m, \Tilde{\alpha}_n]= m \delta_{m,-n}
\end{equation}
In the usual fashion, $\alpha_m$ and $\Tilde{\alpha}_m$ are interpreted as left- and right- moving operators. For $m<0$ we have creation operators while for $m>0$ annihilation operators. Following [ref28 in paper] we can write the Virasoro generators $L_0, L_{\pm 1}$ as 
\begin{equation}
    L_m = \frac{1}{2} \sum_{n =- \infty}^{\infty} : \alpha_{m-n} \alpha_n :
\end{equation}
which obey the algebra 
\begin{equation}
    [L_{\pm 1},L_0]= \pm L_{\pm 1}, \quad \quad [L_{1}, L_{-1}]= 2L_0
\end{equation}
the same is for $\Tilde{L}_0, \Tilde{L}_{\pm1}$. \\
 There are two parameter family of vacua distinguished by their eigenvalues of $\alpha_0$ and $\Tilde{\alpha}_0$ i.e the momentum and winding mode of each vacuum. This is equivalent of labeling the vacua by their eigenvalues $h$ and $\Tilde{h}$ of the Virasoro generators $L_0$ and $\Tilde{L}_0$.
 \begin{equation}
     h= \frac{1}{2} \Big( p + \frac{Rw}{2}\Big)^2, \quad \quad  \Tilde{h}= \frac{1}{2} \Big( p - \frac{Rw}{2}\Big)^2,
 \end{equation}
 where we denote the vacuum state as $|\ket{0;h,\Tilde{h}}$. 
We will show that the only de Sitter invariant vacuum is the $p=w=0$ vacuum i,e $|\ket{0;0,0}$. \\
Excited states are created by acting on the vacuum with creation operators $ \alpha_m$ and $\Tilde{\alpha}_m$ for $m<0$, where a general state will be labeled as $|\ket{n,\Tilde{n}; h, \Tilde{h}}$ where $n$ and $\Tilde{n}$ are eigenvalues of $L_0-h$ and $\Tilde{L}_0 - \Tilde{h}$, respectively.\\
Introducing the operator 
\begin{equation}
    H= L_0 + \Tilde{L}_0
\end{equation}
which generates translation in $\tau$. Even that H is not thought as de Sitter Hamiltonian, it agrees with the flux of de Sitter stress-energy through the sphere at $\tau=0$. The action of $H$ on a general state $|\ket{n,\Tilde{n}; h, \Tilde{h}}$ is 
\begin{equation}
\begin{split}
    H |\ket{n,\Tilde{n}; h, \Tilde{h}} =  &\bigg( h \Tilde{h} + n + \Tilde{n} \bigg) |\ket{n,\Tilde{n}; h, \Tilde{h}} \\
    & =\bigg( p+ \frac{R^2 w^2}{4} + n + \Tilde{n} \bigg) |\ket{n,\Tilde{n}; h, \Tilde{h}} 
    \end{split}
\end{equation}
where the energy of a such a state is $E= n+\Tilde{n}+ h +\Tilde{h}$.\\
The $1+1$ de Sitter space-time has three independent killing vector fields. The Killing fields act on $\mathcal{H}_{aux}$ via the operators $J,B_1$ and $B_2$ which satisfy. the $SO(2,1)$ algebra
\begin{equation}
    [B_1,B_2] = iJ, \quad [B_1,J]=iB_2, \quad [B_2,J]=-iB_1
\end{equation}
One can express the $SO(2,1)$ generators in terms of Virasoro generators via
\begin{equation} \label{generators}
    \begin{split}
        J&= L_0 - \Tilde{L}_0 \\
        B_1 &= \frac{1}{2} \Big(L_1 + L_{-1} + \Tilde{L}_1 + \Tilde{L}_{-1} \Big) \\
        B_2 &= -\frac{i}{2} \Big(L_1 + L_{-1} - \Tilde{L}_1 + \Tilde{L}_{-1} \Big)
    \end{split}
\end{equation}

Because boost generators $B_1$ and $B_2$ contain both raising and lowering Virasoro generators it is hard o construct a state that is boost invariant. What can be checked is that the only de Sitter invariant state is $p=w=0$ vacuum $|\ket{0;0,0}$ where $|\ket{0;0,0} \in \mathcal{H}_{aux}$.

\section{Group averaging for SO(2,1)}
We use group averaging over de Sitter group SO(2,1) to construct physical states that satisfy $M_{AB} \ket{\psi}=0$ where $M_{AB}$ are generators of the group. For $SO(2,1)$ the condition $M_{AB} \ket{\psi}=0$ reads \eqref{generatorconditioon}. These physical states live in the physical Hilbert space $\mathcal{H}_{phys}$ but first we define the space $\mathcal{H}_{bas}= \{ |\ket{\psi}_{phys} \}$ such that $|\ket{\psi}_{phys} $: 
\begin{equation} 
    \begin{split}
        &J |\ket{\psi}_{bas} \\
        & L_1|\ket{\psi}_{bas} = \Tilde{L}_1 |\ket{\psi}_{bas}=0
    \end{split}
\end{equation}
and which are the states in the subspace corresponding  to eigenvalues $E =n+\Tilde{n}+ h +\Tilde{h} >1$
\begin{equation}\label{generatorconditioon}
    \begin{split}
      &  J |\ket{\psi}_{bas}=0w \\
      & B_1 |\ket{\psi}_{bas} = (L_1 + \Tilde{L}_1) |\ket{\psi}_{bas}, \quad \quad \bra{\psi}_{bas} |_{phys} B_1 = \bra{\psi}_{phys}| (L_1 + \Tilde{L}_1 )
    \end{split}
\end{equation}
We can decompose the group elements of $SO(2,1)$ as a product of group elements of two $SO(2)$ rotations and a boost \cite{Barut:1986dd}.
\begin{equation}
   U(g) = e^{i \alpha J} e^{ i\lambda B_1} e^{i \gamma J} 
\end{equation}
where $e^{i \alpha J}$ is the $SO(2)$ rotation through an angle $\alpha$ and $e^{ i\lambda B_1}$ is the boost along one of the Killing vector fields.\\
The Haar measure can be written as 
\begin{equation}
    dg =\frac{1}{4 \pi} d \alpha d \beta d \lambda \sinh{\lambda}
\end{equation}
Then the group averaging inner product of to basic states $\bra{\psi}_{bas}$ is 
\begin{equation} \label{groupaveraging.inner}
     \bra{\psi_1} \ket{\psi_2} = \frac{1}{4 \pi} \int d \alpha d \beta d \lambda \sinh{\lambda} \bra{\psi_1}|e^{i \alpha J} e^{ i\lambda B_1} e^{i \gamma J} |\ket{\psi_2} 
\end{equation}
after some calculations \cite{Marolf_2009} the last integral is 
\begin{equation} \label{innerprod}
    \bra{\psi_1} \ket{\psi_2}= \delta_{\psi_1, \psi_2}
\end{equation}
The equation\eqref{innerprod} completes the  group averaging inner product of two basic states.

The question that we want to answer in this note is, how to relate this calculation with BRST quantization?

\section{The BRST interpretation}
The question that we want to answer in this note is, how to relate this calculation with BRST quantization?\\
A good part of this section is based on Appendix B of \cite{Chandrasekaran:2022cip}\\.
 Let $t_p, p=1,..., dimG_{dS}$ be the linear operators that generate the action of $G_{dS}$ on $ \mathcal{H}_0$. They obey the commutation relations
\begin{equation}
	[t_p,t_q]= f_{pq}^r t_r
\end{equation}
where $f_{pq}^r$ are the structure constants of $G_{dS}$ and obey the Jacobi identity.\\
Next we introduce fermionic operators $c^r$ and $b_s$, known as ghosts and anti-ghosts, respectively, with anti-commutation relations
\begin{equation}
	\begin{split}
		& \{c^r,b_s\}= \delta_s^r \\
		& \{c^r,c^s\}= \{b_r,b_s\}=0
	\end{split}
\end{equation}
Ghost number is defined so that $c^r$ has ghost number 1 and $b_s$ has a ghost number -1. The anti-ghost operators have an irreducible representation on a finite-dimensional vector space $\mathcal{K}$ that contains a state $ \ket{\downarrow}$ of a minimum ghost number, with $ b_s \ket{\downarrow}=0$. Other states in  $\mathcal{K}$ are obtained by acting on $\ket{\downarrow}$ with a polynomial in the c's, while $\mathcal{K}$ contains the state $\ket{\uparrow}= c^1 c^2... c^{dim G_{dS}} \ket{\downarrow}$ of a maximum ghost number \\
The group $G_{dS}$ acts on the combined space $\mathcal{H}_0 \otimes \mathcal{K}$ with generators
\begin{equation}
	T_r = t_r - \sum_{s,t} f_{rs}^t c^s b_t
\end{equation}
and the BRST operator is given by 
\begin{equation}
	Q= \sum_{r} c^r t_r -\frac{1}{2} \sum_{r,s,t} f_{rs}^t c^s b_t
\end{equation}
The cohomology of Q at ghost number  $n$ is defined  as the space of states $\Psi$ of ghost number $n$ that satisfy $Q \Psi =0$ modulo those of the form $   \Psi= Q \chi$. We denote this  cohommology group  as $H^n (Q,\mathcal{H}_0)$\\
In the BRST approach to quantization, the space of physical states is defined via the cohomology of the BRST operator Q at a specific value of the ghost number. In the example of the de Sitter space with weakly coupled gravity, there are two important values of the ghost number, namely the minimum and the maximum possible values.
A state in $\mathcal{H}_0 \otimes \mathcal{K}$ of the minimum possible ghost number is of the general form $ \hat{\Psi} = \Psi \otimes \ket{\downarrow}$ with $\Psi \in \mathcal{H}_0$. The condition $Q \hat{\Psi}=0$ reduced to $ t_a \Psi=0$ which means that $ \Psi$ must be $G_{dS}$ invariant. The problem here is that, except $\Psi_{dS}$ there are no $G_{dS}$ invariant states in $\mathcal{H}_0$.\\
On the other hand if you consider states with maximum ghost number, such states are $\hat{\Psi}= \Psi \otimes \ket{\uparrow}$, then the condition $ Q \hat{\Psi}=0$ is trivial since the ghost number of $ Q\hat{\Psi}$ exceeds the maximum possible value. However, one needs to consider the equivalence relation $\hat{\Psi} \equiv \Psi+ Q \chi$ where $ \chi =\sum_{a} \chi^a b_a \ket{\uparrow}$ for some state $\chi^a \in \mathcal{H}_0$. Then the equivalence relation becomes
\begin{equation}
	\hat{\Psi} \equiv \Psi \sum_{a} t_a \chi^a
\end{equation}
Thus, the conclusion is that for states of maximum ghost number, the equivalence relation reduced to an equivalence relation on $\mathcal{H}_0$, namely $t_a \chi \equiv =0$ for any state $\chi \in \mathcal{H}_0$. This is the same as the equivalence relation $(1-g)\chi \equiv 0$ that you get from group averaging. The next question is how to construct the inner product in the BRST language?\\
There are different ways of quantizing a system, first class of Dirac method,refined algebraic quantization( group averaging) and BRST-BFV quantization method. The Dirac method is based on imposing imposing the constraint conditions on physical states but introducing inner product in this case could be challenging.  Meanwhile in the refined algebraic method instead of imposing constrains on physical states, one modifies the inner product due to constrains. \\
Let us try to build up the inner product between two states in BRST-BFV quantization method.\\
Following \cite{paper3} and \cite{Hwang:1988ak}  the BRST operator Q can be decomposed as
\begin{equation} \label{BRSToperator}
Q = \delta + \delta^\dagger, \quad \delta^2=0, \quad [\delta, \delta^\dagger] =0
\end{equation}
where 
\begin{equation} \label{deltacond}
    \delta \ket{\psi}= \delta^\dagger \ket{\psi}=0
\end{equation}.
with $\ket{\psi}$ being a  physical state. This decomposition is possible  provided that there is a bigrading of the state of space $\Omega$ such that 
\begin{equation} \label{fketfdagerket}
    F \ket{k,m}=k \ket{k,m} \quad \quad F^\dagger = m \ket{k,m}
\end{equation}
where $F$ satisfies 
\begin{equation}
    \delta= [F,Q] \quad \quad N=F-F^\dagger
\end{equation}
with $N$ being the ghost number operator $N= \frac{1}{2}(c^aP_a- c_aP^a)$. The state $\ket{k,m}$ in \eqref{fketfdagerket} represents a general state as a combination of matter states and ghosts.  The operator that makes the grading possible is the ghost number operator which does the grading of the stace space $\Omega$ into $\Omega =  \sum_n \bigoplus \Omega_n$, consequently this does the bigrading for us. 
In \cite{paper1,paper2,paper3} it was shown that there exists a $\delta$-operator which satisfies the \eqref{BRSToperator}. With the construction of $\delta$-operator it is possible to solve \eqref{deltacond} where you get the solution
\begin{equation}
    \ket{\psi}= e^{[Q,\xi]} \ket{\phi}
\end{equation}
where $ \xi$ is fermionic gauge fixing operator with ghost number $-1$ and  $\ket{\phi}$ is a BRST invariant state. If $U$ is an unitary operator, then we can add 
\begin{equation}
    \delta^\prime = U \delta U^\dagger, \quad \quad [Q,U]=0
\end{equation}
to the conditions \label{BRSTopc}. Then for a physical state $\ket{\psi}$ we have
\begin{equation}
    \ket{\psi^\prime}= U \ket{\psi}
\end{equation}
and BRST invariant states are determined up to a unitary transformation
\begin{equation}
    \ket{\psi^\prime}= e^{[Q,\xi^\prime]} \ket{\psi^\prime}, \quad \psi^\prime = U \psi U^\dagger, \quad \quad \ket{\phi^\prime} = U \ket{\phi}
\end{equation}
If $U$ is of the form $U=e^{[Q,\xi]}$, then it represents unitary gauge transformation. \\
Let $t_a$ be the generators of the gauge group which are bosonic and satisfy the Lie Algebra relation
\begin{equation}
[t_a,t_b]= i f_{ab}^c t_c
\end{equation}.

and the BRST operator is 
\begin{equation}
    Q= t_a c^a- \frac{1}{2} i f_{bc}^a P_a c^b c^c - \frac{1}{2} i f_{ab}^b c^a - \Bar{P}_a \pi^a
\end{equation}
where $c^a, b_a$ are ghost and anti-ghosts respectively, $P_a, \Bar{P}^a$ are their conjugate momenta, and $\pi_a$ are the conjugate momenta to the Lagrange multipliers $v^a$ which satisfy the algebra 
\begin{equation}
    [c^a,P_b]=[b^a,\Bar{P}_b] \delta_b^a, \quad \quad [v^a, \pi_b] = \delta_b^a
\end{equation}
A physical state could be written as 
\begin{equation} \label{psisol}
\ket{\psi}= e^{\alpha [\rho,Q]} \ket{\phi}
\end{equation}
From \cite{paper1} it is found that 
\begin{equation}
    \rho= P_a v^a
\end{equation}
and the fact that $\ket{\phi}$ satisfies the conditions
\begin{equation}
    c^a \ket{\phi}= b_a \ket{\phi} =\pi_a \ket{\phi}
\end{equation}
which means that $\ket{\phi}$ has no ghost and Lagrange multipliers dependence.\\
We can write the inner product for solutions which satisfy \eqref{psisol} as 
\begin{equation}
   \braket{\psi|\psi}= \bra{\phi} e^{2 \alpha [\rho,Q]} \ket{\phi}
\end{equation}
and for the gauge fixing we have 
\begin{equation}
\ket{\psi}_{\alpha} =U(\beta) \ket{\psi}_{\pm} \quad \quad \ket{\psi} \equiv \ket{\psi}
\end{equation}
where $\alpha= \pm e^\beta$ and $U(\beta)$ is the unitary operator
\begin{equation}
    U(\beta)= e^{-i \beta[v^a b_a,Q]}
\end{equation}
In the case of $SL(2,R)$ gauge theory the structure constant is given by $\epsilon_{ab}^c$ and the gauge fixing term is the same as above.\\
 Following the inner product $\bra{\phi} e^{[\rho,Q]} \ket{\phi}$ can be written as 
 \begin{equation} \label{innerproductbrst}
    _{\pm}\braket{\psi|\psi}_{\pm}= \bra{\phi} e^{[\rho,Q]} \ket{\phi}=
     \begin{cases}
     \bra{\phi} e^{\psi_a^\prime v^a} e^{i (L^{-1})_b^a (iv) P_A \Bar{P}^b} \ket{\phi} \\
     \bra{\phi} e^{i (L^{\dagger -1})_b^a (iv) P_A \Bar{P}^b} e^{\psi_a^{ \prime \prime} v^a} \ket{\phi}
     \end{cases}
 \end{equation}
where $L_b^a(iv)$ is the left invariant vielbein on group manifold and $\psi^\prime = \psi_a - \frac{1}{2} i f_{ab}^c$ and $\psi^{\prime \prime}= (\psi^\prime)^\dagger$.\\
For $\ket{\phi}$ we can write 
\begin{equation}
    \ket{\phi}=\ket{\phi_1}\ket{0}_{\pi} \ket{0}_{cb}
\end{equation}
where $\ket{0}_{\pi}$ is a Lagrange multiplier vacuum and $\ket{0}_{cb}$ is a ghost vacuum. The ghost vacuum is normalized as \cite{paper7}
\begin{equation}
   _{cb}\bra{0} (i P_a \Bar{P}^a \ket{0}_{cb}=1
\end{equation}
In the case of $SL(2,R)$ gauge theory
\begin{equation}
    [\rho,Q]_{+}= \psi_a v^a +\frac{1}{2} i \epsilon_{bc}^c ( P_c c^b -c^b P_c) v^a + i P_a \Bar{P}^a
\end{equation}
Using equation \eqref{innerproductbrst} we can write 
\begin{equation} \label{++--}
    \begin{split}
     &   _{+}\braket{\psi | \psi}_{+}= \bra{\phi} e^{[\rho,Q]_+} \ket{\phi}= \bra{\phi_1}_{\pi} \bra{0} e^{\psi_a v^a}det(L^{-1})_b^a (iv) \ket{0}_{\pi} \ket{\phi_1} \\
     &_{-}\braket{\psi | \psi}_{-}= \bra{\phi} e^{[\rho,Q]_+} \ket{\phi}= \bra{\phi_1}_{\pi} \bra{0} e^{-\psi_a v^a}det(L^{-1})_b^a (iv) \ket{0}_{\pi} \ket{\phi_1}
    \end{split}
\end{equation}
The explicit form of $det(L^{-1})^a_b(iv)$ is given as
\begin{equation} \label{det}
    det(L^{-1})^a_b(iv)=  \frac{2(1-\cos v)}{v^2}
\end{equation}
substituting \eqref{det} in \eqref{++--} then we get
\begin{equation} \label{finalinner}
    \begin{split}
        &  _{+}\braket{\psi | \psi}_{+}= \int d^3 u \frac{2(\cosh u -1)}{u^2} \bra{\phi_1} e^{i \psi_a v^a} \ket{\phi} \\
        & _{-}\braket{\psi | \psi}_{-} - = _{+}\braket{\psi | \psi}_{+}
    \end{split}
\end{equation}
Finally we need to factorize the exponential $e^{\psi_a v^a}$ as 
\begin{equation} \label{decomp}
    e^{\psi_a v^a}= e^{\alpha \psi_3} e^{\beta \psi_2} e^{\gamma \psi_3} = e^{\alpha \psi_3} e^{\lambda \psi_2} e^{\gamma \psi_3}
\end{equation}
then for the case of $SL(2,R)$ which is isomorphic with $SO(2,1)$ we can write the decomposition as in \eqref{decomp}. Then for $d^3 u \frac{2(\cosh u -1)}{u^2}$ we can write 
\begin{equation} \label{measure}
d^3 u \frac{2(\cosh u -1)}{u^2}= i d\alpha d\lambda d\gamma \sinh \lambda
\end{equation}
Substituting \eqref{measure}  in \eqref{finalinner} we end up 
\begin{equation} \label{innerproduct.int}
    _{+}\braket{\psi | \psi}_{+}= \int d\alpha d\lambda d \gamma \bra{\phi_1} e^{\alpha \psi_3} e^{\lambda \psi_2} e^{\gamma \psi_3} \ket{\phi_1}
\end{equation}

Identifying $\psi_3=J$ and $ \psi_2=B_1$ we see that the integral.
\eqref{innerproduct.int} is similar to \eqref{groupaveraging.inner}.

\section{Discussion}
Finally, I would mention a paper written by Witten \cite{Witten:2022xxp} where the canonical formalism for gravity has been studied and a gauge fixing model has been constructed in the case of the Hilbert space of quantum gravity for asymptotically Anti de Sitter space-time.
Einstein's equations are equations for a metric $h$ on an initial value surface and symmetric tensor field $K_{ij}$ on that surface. For a general relativity theory with a cosmological constant $\Lambda$ and no matter fields, equations read 
\begin{equation} \label{WheelerdeWitt}
    \begin{split}
        &P^j(\Vec{x}) = 0 \\
        & \mathcal{H}(\Vec{x})=0
    \end{split}
\end{equation}
where the first equation is known as momentum constrain and the second one as Hamiltonian constrain. Both together they are known as Wheeler-DeWitt equation.\\
In the most basic version of the canonical approach of quantum gravity, the wave function $\Psi(h)$ is a function of the metric of S and the "usual" (traditional) interpretation of Wheeler-DeWitt equation is that a quantum state has to be taken from the function $\Psi(h)$ satisfying Wheeler-DeWitt equation.\\
As explained in a paper by Witten \cite{Witten:2022xxp}, one can construct such a states but there are drawbacks. The idea of constructing a Hilbert space of quantum states is that one wants to factorize transition amplitudes and sum over intermediate states but the difficulty is that there is no argument that such a wave function takes part in the “sum over states” formulas and another drawback is that the path integral which is supposed to be computed by such a wave function is not well defined in perturbation theory \cite{Witten:2018lgb,Avramidi:1997sh,Anderson:2006lqb}.\\
Recently, in two papers \cite{Chakraborty:2023yed,Chakraborty:2023los} it was demonstrated that by solving the WdW equation on a late-time slice in the non-gravitational limit, one can obtain the states and inner product of group averaging \cite{Higuchi:1991tk,Higuchi:1991tm}.
As we mentioned above, this is the "usual" (traditional) interpretation of Wheeler-DeWitt equation, however there is also a “revised” description which one constructs states that could appear in the factorization formulas, sometimes known as refined algebraic quantization (RAQ) or group averaging. In this method one still imposes the momentum constraint to the wave function $P^i \Psi(h)=0$ but the Hamiltonian constraint $\mathcal{H} \Psi=0$ is replaced by an equivalence relation. 
\begin{equation} \label{equivalrel}
    \Psi(h)= \Psi(h) + \sum \mathcal{H}(\Vec{x}_i) \xi_i(h)
\end{equation}
In this description, any state that satisfies the momentum constraint is considered physical and the inner product of two such states is given in a similar form as \eqref{integral1}.\\
The "usual"(traditional) and the "revised" Wheeler-DeWitt theories can be viewed as special cases of a BRST quantization. In BRST quantization, one introduces ghosts fields with ghost number 1 which transform as generators of gauge group, while in the case of gravity ghost fields are anti-commuting vector fields $c^\mu(\Vec{x},t)$. In the paper by Witten \cite{Witten:2022xxp} the BRST operator in the case of gravity has been constructed. This operator obeys the usual BRST operator relations, $Q^2=0$ which defines the cohomology of states $\Psi$ that satisfy $Q \Psi=0$. In the case where we are interested in the states that are annihilated by $c^0$ and not by $c^i$ the condition $Q \Psi=0$ gives the momentum constraint$ \mathcal{H} \Psi=0$  and for the Hamiltonian constraint one should get \eqref{equivalrel}, whereas if one is interested in the states that are annihilated by all modes of $c^\mu$ then for the Hamiltonian constraint gets $\mathcal{H} \Psi=0$.\\
As it is explained in the paper \cite{Witten:2022xxp} neither of these methods correspond to the usual BRST quantization method where one introduces ghosts and anti-ghosts and where those two fields are treated similarly as other fields.\\
It would be interesting to further study \cite{Witten:2022xxp} and analyse the gauge-fixing procedure, since it was mentioned that the usual treatment of the BRST quantization does not correspond to "revised" Wheeler-DeWitt theory;group averaging, and the task would be obtain \eqref{innerproduct.int} from a gauge-fixing constructed following \cite{Witten:2022xxp}.
\appendix

\acknowledgments
I’m grateful to Edward Witten for constructive criticism and comments on the first draft of the paper. I also would like to thank Zeqë Tolaj for useful comments and discussions.
I would like to thank Atsushi Higuchi for point out to me the references \cite{Moncrief:1978te,Moncrief:1979bg,Hajicek:1994px} and for pointing out a more mathematical treatment known as  Riefel induction \cite{Landsman:1993xe} and Joydeep Chakravarty for useful comments on their method of obtaining group averaged Hilbert space by solving WdW equation.

\nocite{Witten:2022xxp,Witten:2018lgb,Witten:2021jzq,Witten:2018lgb}
\nocite{Shvedov:2001ai}
\nocite{Park:2015xoa,Nurmagambetov:2018het}
\bibliographystyle{JHEP}
\bibliography{references.bib}
 \end{document}